\documentstyle[preprint,aps]{revtex}
\begin{document}
%\twocolumn[\hsize\textwidth\columnwidth\hsize\csname@twocolumnfalse\endcsname

\title{Variational calculation of 
positronium-helium-atom scattering length}

\author{Sadhan K. Adhikari}
\address{Instituto de F\'{\i}sica Te\'orica, 
Universidade Estadual Paulista,\\
01.405-900 S\~ao Paulo, S\~ao Paulo, Brazil\\}

\date{\today}
\maketitle

\begin{abstract}

We present a variational basis-set calculational scheme for elastic
scattering of positronium atom by helium atom in S wave and apply 
it to the calculation of the scattering length.  Highly
correlated trial functions with appropriate symmetry are used in this
calculation. We report numerical result for the scattering length in
atomic unit: $(1.0\pm 0.1)a_0$. This corresponds to a zero-energy
elastic cross
section of $(4.0\pm 0.8)\pi a_0^2$.

{\bf PACS Number(s): 34.90.+q, 36.10.Dr}

\end{abstract}

%\vskip1.5pc]

\newpage
\section{Introduction}

Recent successful measurements of ortho positronium (Ps) scattering 
cross sections  by
H$_2$, N$_2$, He, Ne, Ar, C$_4$H$_{10}$, and C$_5$H$_{12}$
\cite{1,2,3,4,4a,4aa,5} have spurred renewed theoretical activity in
this
subject 
 \cite{6,7,8,9,10,10a}.  Of these, the
Ps-He system is of special interest as it is the simplest system in which
there are experimental results for total cross section \cite{2,3,4} and
pickoff quenching rate \cite{4a,11}.  The experimental results for partial
and
differential cross sections for this system should be available soon
\cite{1}.
A complete
understanding of this system is necessary before a venture to more complex
targets. 

The pioneering calculations in this system using the static exchange
approximation were performed by Barker and Bransden \cite{12,13} and by
Fraser and Kraidy \cite{14,15}.  There have also been R-matrix \cite{6},
close-coupling (CC) \cite{9,10} and model-potential \cite{16} calculations
for Ps-He scattering. 
More recently, there has been  successful
calculation of Ps scattering by H \cite{17}, He \cite{18,19,20}, Ne
\cite{20}, Ar
\cite{20}, and H$_2$ \cite{21}
using a regularized model exchange potential in a
coupled-channel formulation.

However, there is considerable discrepancy among the different
theoretical Ps-He cross sections at zero energy which we discuss below. 
The static-exchange calculation 
by Sarkar and Ghosh \cite{9}, and by
Blackwood {\it et al.} \cite{6}  yielded 
  $14.38\pi a_0^2$ (at 0.068 eV), and $14.58\pi a_0^2$ (at 0 eV), 
respectively, for
the elastic cross section. The inclusion of more states of Ps in the CC 
\cite{10} and R-matrix \cite{6} calculations does not change these results
substantially. The pioneering static-exchange calculations by Barker and 
Bransden \cite{12}  yielded 13.04 $\pi a_0^2$ and 
by Fraser \cite{13} yielded  $14.2 \pi a_0^2$ for 
zero-energy Ps-He cross section. 
These results are in good agreement with each other. However,  the
model potential calculation by Drachman and Houston \cite{16} yielded
$7.73\pi a_0^2$ and by this author \cite{19} yielded $3.34\pi a_0^2$ for the
zero-energy Ps-He cross section. So there is considerable discrepancy in
the results of different theoretical calculation of low-energy Ps-He
elastic scattering.

On the experimental front, there have been conflicting results for the
low-energy Ps-He elastic cross section by Nagashima {\it et al.} \cite{4},
who measured a cross section of $(13\pm 4)\pi a_0^ 2$ at 0.15 eV, by
Coleman {\it et al.} \cite{4aa}, who reported $9\pi a_0^ 2$ at 0 eV, by
Canter {\it et al.} \cite{4a}, who found $8.47\pi a_0^ 2$ at 0 eV, and by
Skalsey {\it et al.} \cite{3}, who measured $(2.6\pm 0.5)\pi a_0^ 2$ at
0.9 eV.  It is unlikely that these findings could be consistent with each
other.

The results for the total cross section of Ps scattering obtained from  
the
coupled-channel
calculation employing the model potential \cite{18,19,20}  are in
agreement with
experiments of Refs. \cite{1,2,3} at low
energies.  For Ps-He, this model, while  agrees \cite{18,19,20} with the
experimental 
total cross sections \cite{1,3} in the energy range 0 to 70 eV, reproduces
\cite{22}
successfully the experimental pickoff quenching rate \cite{4a,11}.  All
other
calculations could not reproduce the general trend of
cross sections of Ps-He scattering in the energy range 0 to 70 eV and
yielded 
a much too small quenching rate at thermal energies \cite{12,14,22}. 
However, the very low-energy elastic cross sections of the model-potential
calculation 
\cite{18,19,20} are at variance with the experiments of Refs.
\cite{4,4a,4aa}.

Pointing at the discrepancy above among  different theoretical and
experimental studies, Blackwood
{\it et al.} \cite{6}   called for a ``fully fledged calculation'' to
resolve the
situation. 
Here we present a  variational basis-set
calculational scheme for low-energy Ps-He scattering in S wave below the
lowest Ps-excitation threshold at 5.1 eV. Using this method we report
numerical result for the scattering length of 
Ps-He using a one parameter uncorrelated He ground-state wave
function \cite{25}. 

We present the  formulation for the variational basis-set calculation in
Sec.
II, the numerical result for Ps-He scattering length in Sec. III and a
summary in Sec. IV.

\section{Formulation}

Because of the existence of three identical fermions (electrons) in the
Ps-He
system, one needs to antisymmetrize the full wave function.
 The position vectors  of the electrons $-$ ${\bf
r_1}$ of Ps,   and ${\bf r_2}$  and ${\bf r_3}$ of He $-$ and positron
(${\bf x}$) measured with respect
to (w.r.t.)
the massive alpha particle  at the origin are shown in Fig. 1. 
In this configuration the wave function for elastic scattering in the
electronic 
doublet state of  
Ps-He is
taken as
\begin{eqnarray}\label{wf}
\psi^1_{\bf k}(1,2,3)= 
\phi(2,3)\eta(1)F_{\bf k}(1,2,3)
\chi(1,2,3)
\end{eqnarray}
where ${\bf k}$ is the incident Ps momentum and  
\begin{eqnarray}
\chi(1,2,3)=\alpha(1)\times \frac{1}{\sqrt
2}[\alpha(2)\beta(3)-\beta(2)\alpha(3)],
\end{eqnarray}
represents the doublet wave function of Ps-He and 
where $\alpha$ denotes spin up state and $\beta$ down and $\eta (1)$
denotes the Ps wave function of electron 1.
The He
ground state   wave function $\phi (2,3)$ and the
scattering
function $F_{\bf k}(1,2,3)$ are 
symmetric under the 
exchange of electrons 2 and 3. The spin function $\chi(1,2,3) $ is
antisymmetric under the same exchange. 
The full antisymmetrization
operator  for the three electrons is $
(1-P_{12}-P_{13})(1-P_{23})$ where $P_{ij}$ is an operator for exchange
in both space and spin of electrons $i$ and $j$. As the
scattering wave function (\ref{wf}) above is already antisymmetrized with
respect to
electrons 2 and 3, the operator $(1-P_{23})$ in the antisymmetrizer is
redundant and 
the relevant antisymmetrizer in this case
is ${\cal A} \equiv (1-P_{12}-P_{13})$.  Hence, the fully
antisymmetric state $\psi^{{\cal A}}_{\bf k}$ of Ps-He scattering is
given by 
\begin{eqnarray}\label{yyt}
\psi^{\cal A}_{\bf k}=
{\cal A}\psi^{1}_{\bf k} 
&=& 
\phi(2,3)\eta(1)F_{\bf k}(1,2,3)
\chi(1,2,3)-
\phi(1,3)\eta(2)F_{\bf k}(2,1,3)
\chi(2,1,3) \nonumber \\&-&
\phi(1,2)\eta(3)F_{\bf k}(3,2,1)
\chi(3,2,1).
\end{eqnarray}

The projection of the
Schr\"odinger equation  on the doublet state $\chi(1,2,3)$ is 
\begin{eqnarray}\label{xxt}
\langle \chi(1,2,3) | (H-E) |\psi^{\cal A}_{\bf k}\rangle  = (H-E)
|\phi(2,3)\eta(1)F_{\bf k}(1,2,3) 
-\phi(1,3)\eta(2)F_{\bf k}(2,1,3)\rangle =0,
\end{eqnarray}
with $H$ the full Ps-He Hamiltonian.
The incident Ps energy $E=6.8 k^ 2$ eV.
Using the identities $\langle 
\chi(1,2,3)|\chi(1,2,3) \rangle =1$ and
$\langle \chi(1,2,3)|\chi(2,1,3) \rangle =$ $
\langle \chi(1,2,3)|\chi(3,2,1) \rangle =1/2$,
we see that  the two terms on the right-hand side of Eq. (\ref{yyt}) give
equivalent
contribution which are combined in Eq. (\ref{xxt}) which is rewritten
as
\begin{eqnarray}\label{zzt}
 (H-E) (1-P_{12})
|\phi(2,3)\eta(1)F_{\bf k}(1,2,3) 
\rangle =0.
\end{eqnarray}
Hence after the spin projection to the doublet state the effective
antisymmetrizer to be used on state (\ref{wf}) is ${\cal A}_1 \equiv
(1-P_{12})$. We shall use this antisymmetrizer in the following and
supress the spin functions.

The full Ps-He Hamiltonian $H$ can be broken in the convenient form as
follows:
$H=H_1+V_1
$ where $H_1$ includes  the full kinetic energy and intracluster
interaction of He
and
Ps for the arrangement shown in Fig.  1 and $V_1$ is the sum of the
intercluster
interaction between He and Ps in the same configuration:
\begin{equation}
 V_1= \biggr[
\frac{2}{x}-\frac{2}{r_1}+\frac{1}{r_{12}}-\frac{1}{\rho_2}
+\frac{1}{r_{13}}-\frac{1}{\rho_3}\biggr].
\end{equation}
We employ 
the position vectors   ${\bf s}_j= ({\bf
x+r}_j)/2$, $\rho_j = {\bf x-r}_j$,  ${\bf r}_{ij}={\bf
r}_i-{\bf r}_j$, $i,j=1,2,3$, $i\ne j$.

The
fully antisymmetric state satisfies the Lippmann-Schwinger equation
\cite{ska}
\begin{equation}\label{2x}
|\psi^{1}_{\bf k}\rangle =|\phi^1_{\bf k}\rangle + G_1 M_1
|\psi^{1}_{\bf k}\rangle,
\end{equation}
\begin{equation}\label{3}
M_1 =V_1{\cal A}_1+(E-H_1)(1-{\cal
A}_1)\equiv  {\cal A}_1V_1+(1-{\cal
A}_1)(E-H_1),
\end{equation}
where the channel Green's function is given by $G_1\equiv (E+i0-H_1)^{-1}$
and the
incident 
wave $|\phi^1_{\bf k}\rangle$ satisfies  $(E-H_1)|\phi^1_{\bf k}\rangle =
0. $ 
We are using atomic units (au) in which
$a_0=e=m=\hbar=1$,
where $e$ ($m$) is the electronic charge (mass) and $a_0$ the Bohr radius.

The properly symmetrized transition  matrix for elastic
scattering is defined by 
$\langle\phi^1_{\bf k} | T^{\cal A} | \phi^1_{\bf k} \rangle =
\langle\phi^1_{\bf k}|V_1|\psi^{\cal A}_{\bf k}\rangle = 
\langle\phi^1_{\bf k}|V_1 {\cal
A}_1|\psi^1_{\bf k}\rangle 
=
\langle\psi^1_{\bf k}|{\cal
A}_1V_1|\phi^1_{\bf k}\rangle $ \cite{ska}.
 A basis-set calculational scheme for the transition matrix can be
obtained from the following expression 
\cite{ska1}
\begin{eqnarray}\label{61}
\langle\phi^1_{\bf k} | T^{\cal A} | \phi^1_{\bf k} \rangle =
\langle\psi^1_{\bf k}|{\cal
A}_1V_1|\phi^1_{\bf k}\rangle+ \langle\phi^1_{\bf k}| {\cal
A}_1V_1|\psi^1_{\bf k}\rangle-
\langle\psi^1_{\bf k}|{\cal A}_ 1 V_1- M_1G_1{\cal A}_ 1 V_1 |\psi^1_{\bf
k}\rangle.
\end{eqnarray}
Using Eq. (\ref{2x}), it can be verified that 
Eq. (\ref{61}) is an identity if exact scattering wave functions
$\psi^1_{\bf 
k}$ are used.
If approximate wave functions are used, expression (\ref{61}) 
is stationary w.r.t. small variations of $|\psi^1_{\bf
k}\rangle$ but not of  $\langle 
\psi^1_{\bf
k}|$. This one-sided variational property emerges because of the lack of
symmetry of the formulation in the presence of explicit antisymmetrization
operator ${\cal A}_1$. However, this variational property can be used to
formulate a basis-set calculational scheme with the following trial
functions \cite{ska1}
\begin{equation}\label{62}
 |\psi^1_{\bf
k}\rangle_t=\sum_{n=1}^N a_n|f_n\rangle, \quad _t\langle
\psi^1_{\bf
k}| = \sum_{m=1}^N b_m \langle f_m|,
\end{equation}
where the suffix $t$ denotes trial and $f_n, n=1,2,...,N,$ are the basis
functions.
Substituting Eq. (\ref{62}) into Eq. (\ref{61}) and using this variational
property  w.r.t.  $| \psi^1_{\bf
k}\rangle$ we obtain \cite{ska1}
\begin{equation}\label{63}
_t\langle
\psi^1_{\bf
k}|=\sum_{m=1}^N  \langle \phi^1 _{\bf k}|
{\cal A}_ 1 V_1
|f_n
\rangle   D_{nm}\langle f_m|,
\end{equation}
 \begin{equation}\label{5} (D^{-1})_{mn}= {\langle f_m |{\cal A}_ 1 V_1-
[{\cal A}_1V_1+(1-{\cal A}_1)(E-H_1)]G_1{\cal A}_ 1 V_1 | f_n \rangle}.
\end{equation} Using the variational form (\ref{63}) and definition
$\langle\phi^1_{\bf k} | T^{\cal A} | \phi^1_{\bf k} \rangle =
\langle\psi^1_{\bf k}|{\cal A}_1V_1|\phi^1_{\bf k}\rangle$ we obtain the
following basis-set calculational scheme for the transition matrix
\begin{equation}\label{4} \langle \phi^1 _{\bf k}|T^{\cal A}|\phi^1_{\bf
k}\rangle_t = \sum_{m,n=1} ^ N { \langle \phi^1 _{\bf k}| {\cal A}_ 1 V_1
|f_n \rangle D_{nm}\langle f_m|{\cal A}_ 1 V_1| \phi^1_{\bf k} \rangle }.
\end{equation} Eqs. (\ref{5}) and (\ref{4}) are also valid in partial-wave
form.

In the present S-wave calculation, the basis functions are taken in the
following form \begin{eqnarray} f_m ({\bf r}_2,{\bf r}_3,\rho_1,{\bf
s}_1)&=&\varphi({\bf r}_3) g_m ({\bf r}_2,\rho_1,{\bf s}_1),\\ \label{8}
g_m ({\bf r}_2,\rho_1,{\bf s}_1)&=& \varphi({\bf r}_2)
\eta(\rho_1)e^{-\delta_m r_2 -\alpha_m\rho_1-\beta_m s_1
-\gamma_m(\rho_2+r_{12})-\mu_m (x+r_1)} \frac{\sin(ks_1)}{ks_1},
\end{eqnarray} where $\delta_m, \alpha_m,\beta_m, \gamma_m, $ and $\mu_m$
are nonlinear variational parameters. The ground-state wave function of the
He atom is taken to be $\phi ({\bf r}_2,{\bf r}_3)=\varphi({\bf r}_2)
\varphi({\bf r}_3) $ with $\varphi({\bf r})=\lambda^{3/2}\exp(-\lambda
r)/\sqrt\pi $ and $\lambda =1.6875$ \cite{25} and
$\eta(\rho)=\exp(-0.5\rho)/\sqrt {8\pi} $ represent the Ps(1s) wave
function. For elastic scattering the direct Born amplitude is zero and the
exchange correlation dominates scattering. To be consistent with this, the
direct terms in the form factors $\langle f_m |{\cal A}_1V_1 |\phi^1_{\bf
k}\rangle$ and $\langle \phi^1 _{\bf k}|{\cal A}_1V_1 |f_n \rangle$ are
zero with the above choice of correlations in the basis functions via
$\gamma_m$ and $\mu_m$. This property follows as the above function is
invariant w.r.t. the interchange of ${\bf x}$ and ${\bf r}_1$ whereas the
remaining part of the integrand in the direct terms changes sign under this
transformation. In Ps-He elastic scattering the electron 2 of He is the
active electron undergoing exchange with the electron 1 of Ps whereas the
electron 3 of He is a passive spectator. In this calculation we include in
Eq. (\ref{8}) correlation between electrons 1 and 2.  Consequently, we deal
with integrals in three vector variables -- ${\bf r}_1, {\bf r}_2$ and
${\bf x}$.  If we also include correlation involving electron 3 we shall
have to deal with integration in four vector variables, which is beyond the
scope of the present study. However, we believe that a meaningful
calculation can be performed only with correlation between the active  
electrons 1 and 2.
Hence, to avoid complication we ignore correlation involving electron 3,
which is expected to lead to correction over the present study.

In   S wave at
zero energy,  $\sin(ks_1)/(ks_1)=1$ in Eq. (\ref{8}); also,
$|\phi^1_p\rangle =\varphi({\bf r}_2)
\varphi({\bf r}_3) \eta({\rho}_1){\sin(ps_1)}/{(ps_1)}$. 
The useful matrix elements  of the present approach are explicitly written
as \cite{ska1}
\begin{eqnarray}\label{7}
\langle \phi^1_p|{\cal A}_1V_1|f_n\rangle
&=& 
- \frac{1}{2\pi}\int 
\varphi({\bf r}_1)
\varphi({\bf r}_3)
\eta({\rho}_2)\frac{\sin(ps_2)}{ps_2} 
[ V_1]
f_n({\bf r}_2,{\bf r}_3,\rho_1,{\bf s}_1) 
d{\bf r}_2
d{\bf r}_3
d{\rho_1}
d{\bf s}_1,\\
&=& - \frac{1}{2\pi}\int 
\varphi({\bf r}_1)
\eta({\rho}_2)\frac{\sin(ps_2)}{ps_2} 
[ {\cal V}_1]
g_n({\bf r}_2,\rho_1,{\bf s}_1) 
d{\bf r}_2
d{\rho_1}
d{\bf s}_1,\\
\label{52}
\langle f_m|{\cal A}_1V_1|\phi^1_p \rangle
&=& - \frac{1}{2\pi}\int 
g_m({\bf r}_1,
\rho_2,{\bf s}_2) 
[{\cal V}_1]
\varphi({\bf r}_2)\eta({\rho}_1)\frac{\sin(ps_1)}{ps_1} 
d{\bf r}_2
d{\rho_1}
d{\bf s}_1,\\
\label{53}
\langle f_m|{\cal A}_1V_1|f_n\rangle
&=& - \frac{1}{4\pi}\int 
g_m ({\bf r}_1,
\rho_2,{\bf s}_2) [{\cal  V}_1]
g_n({\bf r}_2,
\rho_1,{\bf s}_1)
d{\bf r}_2
d{\rho_1}
d{\bf s}_1,
\end{eqnarray}
with 
\begin{equation}
{\cal  V}_1= \biggr[
h(x)-h({r_1})+\frac{1}{r_{12}}-\frac{1}{\rho_2}
\biggr], \quad \quad h(x)= \frac{1}{x}+\frac{\exp(-2\lambda x)}{x}
+\lambda \exp(-2\lambda x),
\end{equation}
\begin{eqnarray}\label{a}
\langle f_m | M_1 G_1  {\cal A}_1V_1| f_n \rangle \approx
-\frac{2}{\pi}
\int_0^\infty dp{\langle f_m|{\cal A}_1V_1|\phi^1_p \rangle \langle
\phi^1_p|{\cal A}_1V_1|f_n\rangle},
\end{eqnarray}
where the so called off-shell  term $ (1-{\cal A}_1)(E-H_1)  $
has been neglected for numerical simplification in this
calculation. This term is
expected to
contribute to refinement over the present calculation. 
In this convention  the on-shell t-matrix
element at zero energy is the scattering length: $a= \langle \phi_0^ 1|
T^ {\cal A}|\phi_0 ^ 1
\rangle.  $

All the matrix elements above can be evaluated by a method presented  in
Ref. \cite{am}. We describe it in the following for  $\langle
\phi_p^1 |{\cal A}_1V_1|f_n\rangle$ of Eq. (\ref{7}).  
By a transformation of variables from $({\bf r}_2, \rho_1,{\bf s}_1)$
to $({\bf s}_1,{\bf s}_2, {\bf x})$ with Jacobian $2^6$  and
separating the
radial and angular
integrations, the form factor (\ref{7})  is given by 
\begin{eqnarray}
\langle
\phi_p^1 |{\cal A}_1V_1|f_n\rangle
&=& -\frac{2^6\lambda^3}{16\pi^3}\int_0^\infty
ds_2 s_2^2
  \frac{\sin(ps_2)}{ps_2}
\int_0^\infty ds_1 s_1^2 e^{-\beta_ns_1}
\int _0^\infty dx x^2 e^{-\mu _ n x} 
\nonumber \\ &\times& \int 
e^{-(ar_1+b\rho_1/2)}
e^{-(cr_2+d\rho_2/2)}
e^{-\gamma_n r_{12}}
[ V_1] d\hat s_1 d\hat s_2 d\hat x, \label{13}
\end{eqnarray}
where $a=\lambda +\mu_n$, $b=2\alpha_n+1$,  $c=\lambda +\delta_n$ and 
$d=2\gamma_n+1$.
Recalling that ${\bf r}_j= 2{\bf s}_j -{\bf x}$, ${\bf r}_{12}=2({\bf
s}_1-{\bf s}_2)$, $\rho_j=2({\bf x}-{\bf s}_j), j=1,2$, 
 we employ  the following expansions of the exponentials
in Eq. (\ref{13}) 
\begin{eqnarray}\label{5x}
e^{-a |2{\bf s -x}|-b|{\bf x-s}|}=
\frac{4\pi}{sx}\sum_{lm}G_l^{(a,b)}(s,x)
Y^*_{lm}(\hat s)Y_{lm}(\hat x),
 \end{eqnarray}
\begin{eqnarray}\label{6x}
h({|2{\bf s -x}|})
{e^{-a|2{\bf s -x}|-b|{\bf x-s}|}}  
=
\frac{4\pi}{sx}\sum_{lm}J_l^{(a,b)}(s,x)
Y^*_{lm}(\hat s)Y_{lm}(\hat x),
 \end{eqnarray}
\begin{eqnarray}\label{7x}
\frac{e^{-a|2{\bf s -x}|-b|{\bf x-s}|}}{|{\bf s -x}|}=
\frac{4\pi}{sx}\sum_{lm}K_l^{(a,b)}(s,x)
Y^*_{lm}(\hat s)Y_{lm}(\hat x),
 \end{eqnarray}
\begin{eqnarray}\label{8x}
\frac{e^{-a|{\bf s_1-s_2}|}}{|{\bf s_1 -s_2}|}=
\frac{4\pi}{s_1s_2}\sum_{lm}A_l^{(a)}(s_1,s_2)
Y^*_{lm}(\hat s_1)Y_{lm}(\hat s_2),
 \end{eqnarray}
\begin{eqnarray}\label{9x}
{e^{-a|{\bf s_1-s_2}|}}=
\frac{4\pi}{s_1s_2}\sum_{lm}B_l^{(a)}(s_1,s_2)
Y^*_{lm}(\hat s_1)Y_{lm}(\hat s_2),
 \end{eqnarray}
where the $Y_{lm}$'s are the usual spherical harmonics. 
Using  Eqs. (\ref{5x}) $-$ (\ref{9x}) in 
Eq. (\ref{13})  we get 
\begin{eqnarray}\label{8z}
\langle
\phi_p^1 |{\cal A}_1V_1|f_n\rangle
&=& -{2^8}\lambda^3\int_0^\infty ds_1 e^{-\beta_n s_1}
\int_0^\infty ds_2 {\sin(ps_2)\over {p
s_2}}\int _0^\infty
dxe^{-\mu_nx}\sum_{l=0}^L(2l+1)
\nonumber \\
&\times&
\biggr[h(x)G_l^{(a,b)}(s_1,x)
G_l^{(c,d)}(s_2,x)
B_l^{(2\gamma_n)}(s_1,s_2)
-
J_l^{(a,b)}(s_1,x)
\nonumber \\
&\times&
G_l^{(c,d)}(s_2,x)
B_l^{(2\gamma_n)}(s_1,s_2)
+
\frac{1}{2}G_l^{(a,b)}(s_1,x)
G_l^{(c,d)}(s_2,x)
\nonumber \\
&\times&
A_l^{(2\gamma_n)}(s_1,s_2)
- \frac{1}{2}G_l^{(a,b)}(s_1,x)
K_l^{(c,d)}(s_2,x)B_l^{(2\gamma_n)}(s_1,s_2)
\biggr].
\end{eqnarray}
where the $l$-sum is truncated at $l=L$.
This procedure  avoids  complicated angular integrations
involving  ${\bf s}_1$,
${\bf s}_2$ and   ${\bf x}$.
The matrix element  takes a simple form requiring    straightforward  
numerical
computation of certain radial integrals only.
The functions $G_l$, $J_l$, $K_l$ etc. are easily calculated using  
Eqs. (\ref{5x}) $-$ (\ref{9x}):
\begin{equation}G_l^{(a,b)}(s,x)=\frac{sx}{2}\int_{-1}^1du P_l(u)
e^{-a |2{\bf s -x}|-b|{\bf x-s}|}, \label{8xz}
\end{equation}
where $P_l(u)$ is the usual Legendre polynomial and $u$ is the cosine of
the angle between ${\bf s}$ and ${\bf x}$.
 The integrals (\ref{52}) and (\ref{53}) can
be evaluated similarly. 
For example 
\begin{eqnarray}\label{8zx}
\langle
f_m |{\cal A}_1V_1|f_n\rangle
&=& -{2^7}\lambda^3\int_0^\infty ds_1 e^{-\beta_n s_1}
\int_0^\infty ds_2 e^{-\beta_m s_2}\int _0^\infty
dxe^{-(\mu_n+\mu_m)x}\sum_{l=0}^L (2l+1)
\nonumber \\
&\times&
\biggr[h(x)G_l^{(e,f)}(s_1,x)
G_l^{(g,h)}(s_2,x)
B_l^{(2\gamma_{mn})}(s_1,s_2)
-
J_l^{(e,f)}(s_1,x)
\nonumber \\
&\times&
G_l^{(g,h)}(s_2,x)
B_l^{(2\gamma_{mn})}(s_1,s_2)+
\frac{1}{2}G_l^{(e,f)}(s_1,x)
G_l^{(g,h)}(s_2,x)
\nonumber \\
& \times&  
A_l^{(2\gamma_{mn})}(s_1,s_2)
 -   \frac{1}{2}G_l^{(e,f)}(s_1,x)
K_l^{(g,h)}(s_2,x)B_l^{(2\gamma_{mn})}(s_1,s_2)
\biggr],
\end{eqnarray}
where $e=\lambda+\delta_m+\mu_n$, $f=2\alpha_n+2\gamma_m+1$,
$g=\lambda +\delta_n+\mu_m$,  $h=2\alpha_m+2\gamma_n+1$ and $\gamma_{mn}
=\gamma_m+\gamma_n$. 

\section{Numerical Result}

We tested the convergence of the integrals by varying the number of
integration points in the $x$, $s_1$ and $s_2$ integrals in
Eqs. (\ref{8z}) 
 and (\ref{8zx}) and 
the $u$ integral in Eq. (\ref{8xz}). 
The $x$ integration was relatively easy
and 20 Gauss-Legendre quadrature points appropriately distributed between
0 and 16 were enough for convergence. In the evaluation of integrals of
type (\ref{8xz})  40 Gauss-Legendre quadrature points were sufficient for
adequate convergence. 
The convergence in the
numerical integration over $s_1$ and $s_2$ was achieved with 300
Gauss-Legendre quadrature points between 0 and 12. The maximum value of
$l$ in the sum in Eqs. (\ref{8z}) and (\ref{8zx}), 
$L$, is
taken to be 7  which is
sufficient for obtaining the convergence with the partial-wave
expansions (\ref{5x}) $-$ (\ref{9x}).

We find that a judicial choice of the parameters in Eq. (\ref{8})  is
needed for convergence. The present method does not provide a bound on the
result. Consequently, the method could lead to a wrong scattering length
if an inappropriate (incomplete)  basis set is chosen.  After some
experimentation we find that for good convergence the nonlinear parameters
$\delta_n$ and $\alpha_n$ should be taken to have both positive and
negative values and $\beta_n$ should have progressively increasing values
till about 1.5.  If no care is taken in choosing the parameters a large number
of functions could be necessary for obtaining convergence. 
The results reported in this work are obtained with the
following parameters for the
functions $f_n, n=1,...,14$: $ \{\delta_n,\alpha_n,\beta_n, \gamma_n,\mu_n
\}\equiv
\{ -0.5,   -0.25,    0.3,    0.01,    0.02\}, $ $ 
\{   -0.7,   -0.25,    0.5,    0.04,    0.02\},$ $ 
\{   -0.7 ,  -0.25,    0.7,    0.03,    0.06\},$ $
\{   -0.4,   -0.1,    0.6,   0.2,    0.2\},$ $ 
\{   -0.2,    0.1,    0.8,    0.2,    0.2\},$ $
\{    0.4,   -0.2,    0.6,    0.3,    0.3\}, $ $
\{   -0.2,   -0.1,    0.7,    0.4,    0.3\}, $ $
\{    0.3,    0.2,    0.8,    0.3,    0.4\}, $ $
\{    0.2,    0.2,    1,    0.4,    0.4\}, $ $
\{    0.3,    0.2,    1.2,    0.5,    0.5\}, $ $
\{   -0.5,    0.2,    1.3,    0.6,    0.6\}, $ $
\{   -0.2,    0.1,    1.4,    0.7,    0.7\},$  $
\{  0.3,    0,    1.5,    0.8,    0.8\},$ $
\{   -0.2,   -0.1,    1.6,    0.9,    0.9\}.$
By employing a suitably chosen set  of the parameters we have kept the number of
functions
to a minimum.

\vskip .4cm {Table I: Ps-He scattering
length in au
for different  $L$ and $N$.} 
\vskip .2cm
\begin{centering}

\begin{tabular} {|c| c| c| c |  c| c|  c |c|}
\hline 
$N$ &  $L=0$    &$L=2$ &$L=3$ & $L=4$&$L=5$  & $ L=6$ &$L=7$
    \\
  \hline
1&2.056  &1.867&1.782  &1.721&1.681  &1.655  & 
1.638\\
3&2.418    &$-1.662$&28.016 &4.151&3.061 &2.839& 2.933  \\
5&$-13.653$    &1.563 & 1.234&1.128&1.135 &1.168&  1.190 \\
6&0.712   &0.982&0.782 &0.637&0.650 &0.769 & 0.878 \\
7&3.372    &0.983 &0.792 &0.694&0.727 &0.824&  0.910 \\
8&$-1.657$    &1.124& 0.944&0.877&0.907 &0.971&  1.023 \\
9&6.283     &0.976 &0.832&0.801&0.841 &0.897  & 0.943\\
10&1.332     &1.123 &0.941 &0.897&0.929  &0.981& 1.023  \\
11&1.225    &1.112 &0.945&0.886&0.909 &0.963  &1.011\\
12&0.756    &1.468 &0.995&0.918&0.931 &0.964   &0.977\\
13&1.229    & 1.197&1.008 &0.944 &0.970 & 1.028 & 1.022\\
14& 1.061   &1.249 &1.060&0.980&0.983 & 1.026& 1.019\\
\hline
\end{tabular}

\end{centering}
\vskip 0.4cm

In Table I we show the convergence pattern of the present calculation
 w.r.t. the number of partial waves $L$ and
basis functions $N$ used in the calculation. 
The convergence is satisfactory considering that we are dealing with a
complicated five-body problem.
 However, as the present
calculation does not
produce a bound on the result, the 
convergence is not monotonic with
increasing $N$. 
The final result of the
present
calculation is that for $N=14$ and $L=7$: $a= 1.02$ au. Although it is
difficult to provide a quantitative measure of convergence, from the 
fluctuation of this result for large $N$ and $L$  we believe the 
error in
our result to be
less than 10$\%$, so that the final Ps-He scattering length
is taken as $a=(1.0 \pm 0.1)$ au. The results for large $N$ and $L$
reported in Table I all lie in this domain.

The maximum number of functions ($N=14$) used in this
calculation is also pretty small, compared to those used in different
Kohn-type variational calculations for electron-hydrogen ($N = 56$) 
\cite{eh}, positron-hydrogen ($N\le 286$)\cite{ph}, and positron-helium
($N\le 502$) \cite{vr} scattering. Because
of the explicit appearance of the Green's function, the present basis-set
approach is similar to the Schwinger variational method.  Using the
Schwinger method, convergent results for electron-hydrogen \cite{ehs} and
positron-hydrogen \cite{phs} scattering have been obtained with a
relatively small basis set ($N\sim 10$). These suggest a more rapid
convergence in these problems with a Schwinger-type method.

\section{Summary and Discussion}

To summarize, we have formulated a basis-set calculational scheme for
S-wave Ps-He elastic scattering below the lowest inelastic threshold
using a variational expression for the transition matrix. We illustrate
the method numerically by calculating the scattering length in the
electronic doublet state: $a= (1.0\pm 0.1) $ au. This corresponds to a
zero-energy cross section of $(4.0\pm 0.8) \pi a_0^2$ in reasonable
agreement with a model calculation by this author ($3.34 \pi a_0^2$)
\cite{19} and the experiment of Skalsey {\it et al.} [$(2.6\pm0.5)\pi
a_0^2$ at 0.9 eV] \cite{3}. This calculation as well as our previous
studies of Ps-He scattering using a model exchange potential \cite{18,20}
possibly consolidate the experimental result of Skalsey {\it et al.}
However, these low-energy Ps-He elastic scattering cross sections are in
disagreement with other experiments by Nagashima {\it et al.} [$(13\pm
4)\pi a_0^2$ at 0.15 eV] \cite{4}, by Canter {\it et al.} ($8.47\pi
a_0^2$) \cite{4a}, and by Coleman {\it et al.} ($9\pi a_0^2$) \cite{4aa},
as well as with conventional static-exchange calculations of Refs.
\cite{6,9,10,12,14} ($\sim 14 \pi a_0^2$) and a model potential
calculation of Ref. \cite{16} ($7.73 \pi a_0^2$).

As the effective interaction for elastic scattering between Ps and He is
repulsive in nature, a smaller scattering length as obtained in this study and
in Refs. \cite{18,19} would imply a weaker effective Ps-He interaction. This
would allow the Ps atom to come closer to He and would lead \cite{22} to a large
pickoff quenching rate and a large $^1Z_{\mbox{eff}}$ $(\sim 0.11)$ in agreement
with experiment \cite{4a,11}.  The conventional close-coupling \cite{10},
R-matrix \cite{6} and static-exchange \cite{9,12,13,14} models yielded a
much too large scattering length corresponding to a stronger repulsion between
Ps and He.  Consequently, these models led to a much too small
$^1Z_{\mbox{eff}}$ $(\sim 0.04)$ \cite{12,13,14,22} in disagreement with
experiment \cite{4a,11}. This is addressed in detail in Ref. \cite{22} where we
established a correlation between the different scattering lengths and the
corresponding $^1Z_{\mbox{eff}}$. This correlation suggests that a small Ps-He
scattering length as in this work is consistent with the large experimental
$^1Z_{\mbox{eff}}$.

Although we have used a simple wave function for He in this
complex five-body 
calculation we do not believe that the use of a more refined He wave
function would substantially change our findings and conclusions.
However, independent  calculations and accurate experiments at
low energies are welcome  for a satisfactory resolution of this
controversy.

The work is supported in part by the Conselho Nacional de Desenvolvimento -
Cient\'\i fico e Tecnol\'ogico,  Funda\c c\~ao de Amparo
\`a Pesquisa do Estado de S\~ao Paulo,  and Finan\-ciadora de Estu\-dos e
Projetos of Brazil.

\vskip 1 cm
\newpage
{\bf Figure Caption:}

Figure number 1. Different  position vectors for the Ps-He system w.r.t.
the
massive alpha
particle
 at the origin in arrangement 1 with electrons 2 and 3 forming He and  1
forming  Ps. The arrows
on the electrons indicate the orientations of spin $-$ up and down.

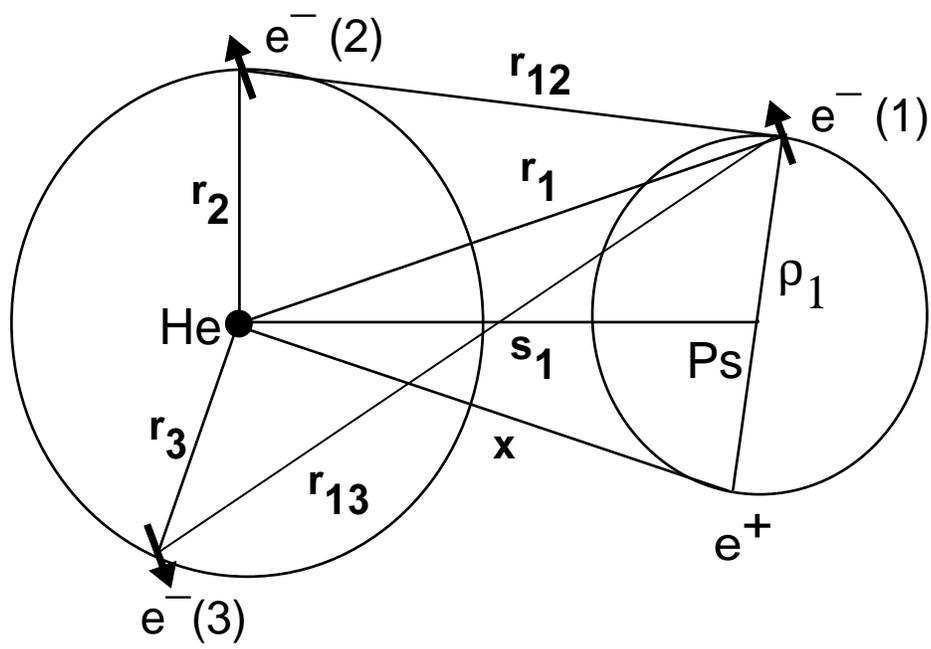
\end{document}